\documentclass[a4paper]{jpconf}
\usepackage{amssymb}
\usepackage{amsthm}
\usepackage{latexsym}
\usepackage{graphicx,epsfig}
\def\be {\begin{equation}}
\def\ee {\end{equation}}
\def\ba {\begin{eqnarray}}
\def\ea {\end{eqnarray}}
\def\nn {\nonumber}
\def\bc {\begin{center}}
\def\ec {\begin{center}}

\def\a  {\alpha}
\def\b  {\beta}
\def\c  {\gamma}

\def\d  {\delta}
\def\D  {\Delta}
\def\e  {\epsilon}

\def\k  {\kappa}

\def\n  {\nu}

\def\O  {\Omega}
\def\p  {\pi}

\def\r  {\rho}

\def\t  {\tau}
\def\la {\label}
\def\le {\left}
\def\ri {\right}
\def\pa {\partial}
\def\f {\frac}

\def\no {\noindent}
\def\bi {\begin{itemize}}
\def\ei {\end{itemize}}

\def\ub {\underbrace}
\def\bc {\begin{center}}
\def\ec {\end{center}}

\newcommand{\bdm}{\begin{displaymath}}
\newcommand{\edm}{\end{displaymath}}

\begin{document}
\title{Entanglement as a source of black hole entropy}
\author{\textbf{Saurya Das}}
\date{}
\address{Department of Physics, \\
University of Lethbridge, \\
4401 University Drive, Lethbridge, Alberta, CANADA T1K 3M4}
\ead{\textbf{saurya.das@uleth.ca }}
\author{\textbf{S. Shankaranarayanan}}
\address{Max-Planck-Institut f\"ur Gravitationphysik,\\
Albert-Einstein-Institut, Am M\"uhlenberg 1, D-14476 Potsdam, Germany} 
\ead{\textbf{shanki@aei.mpg.de}}

\begin{abstract}
We review aspects of black hole thermodynamics, and show how
entanglement of a quantum field between the inside and outside of a
horizon can account for the area-proportionality of black hole
entropy, provided the field is in its ground state.  We show that the
result continues to hold for Coherent States and Squeezed States,
while for Excited States, the entropy scales as a power of area less
than unity.  We also identify location of the degrees of freedom which
give rise to the above entropy.
\end{abstract}

\section{Introduction}

The field of black hole thermodynamics started around the time 
Bekenstein and Hawking argued that black holes have 
entropy and temperature, given respectively by \cite{bekhaw}:
\be
S_{BH} = \f{A_H}{4\ell_{Pl}^2}~~\mbox{and}~~  
T_H = \f{\hbar c^3}{8\p GM} ~, 
\la{temp1}
\ee
where $M$ and $A_H$ are the mass and horizon area of the black hole, 
and  
${\ell_{Pl}}=\sqrt{G\hbar/c^3}$ is the Planck length in four dimensions.  
Together, they satisfy the following laws, which are
analogous to the laws of classical thermodynamics
\footnote{In subsequent formulae, we will assume $c=1=\hbar$.}
:
\ba
d(Mc^2) &=& T_HdS + \Phi dQ + \O dJ ~~~\mbox{(First law)} \\
\D S_{BH} &\geq&  0~~~ \mbox{(Second law)}~.
\ea
The question naturally arises as to whether there are miscroscopic, elementary
states, consistent with the black hole
temperature and other macroscopic parameters,
whose logarithm gives its entropy:
\be
S_{BH} = \ln\O~,
\ee
and if so, then what those states are. Another related question
in this area of research is the following: 
is the area proportionality of the entropy, also
know as the {\it Area Law} (AL) in Eq.(\ref{temp1}) 
(as opposed to volume proportionality, which is the case for normal
thermodynamic systems) 
more generic, or in fact universal, beyond the realm of black 
holes? This has given rise to the so-called 
{\it Holographic Hypothesis}, which conjectures that the
information content of our universe may be encoded in a 
two dimensional screen which surrounds three dimensional space. 

Finally, the expectation that black holes, after absorbing 
matter (and information contained therein) will radiate itself away 
in the form of pure thermal radiation at the temperature (\ref{temp1}),
has given rise to the {\it Information Loss Problem}: thermal 
radiation does not contain any information, except that of the
temperature of the emitting source, so if the black hole 
disappears completely, does the information do so as well? 
While there is nothing in principle which prevents this
form happening, it requires the quantum mechanics which governs
the evolution to be non-unitary, since unitary 
evolution takes pure states to pure states.  
This eventually raises the question:
{\it does quantum mechanics break down in the presence of 
a black hole?}
%
%
To be able to address one or more of the above questions, 
one has to first explore origins of Black Hole Entropy and temperature. 
Among the various approaches that have been undertaken, the two
most popular are: String Theory and Loop Quantum Gravity.
In the first, degeneracy of strings attached to $D$-branes are
claimed to give rise to the required degeneracy, while in the
second, the spin-network states at the horizon are invoked
for the same purpose. There have been some other interesting 
approaches as well \cite{stringsetc,yarom}. 

In addition to the above, there has been another idea, which goes as
follows: consider a free scalar field, propagating in a spacetime
containing a black hole. Now, do a partial tracing over the degrees of
freedom (DOF) of the field which are inside the horizon. The resultant
density matrix is mixed, due to the information being `hidden' by the
horizon (although the entire system, before tracing, is pure). When
one computes the {\it Entanglement Entropy} associated with this, the
latter turns out to be proportional to the horizon area
\cite{bkls,sred}! [The analytical proof for the area law have recently
been shown in Refs. \cite{eisert}.] It is this approach which we will
follow in this article.  Here we will critically examine an important
assumption that the authors had made in their analyses: that the
scalar field is in its {\it Ground State} (GS). While this simplifying
assumption is justified and even natural to make in a first
calculation, we would like to see how important it really is to
validity of the AL.  Indeed we will find that while replacing the GS
by generic {\it Coherent State} (CS) or a class of {\it Squeezed
States} (SS) does not affect the AL, doing so with a class of {\it
Excited States} (ES) does change the AL, and in fact renders the
entropy proportional to a power of area less than unity.

One can ask as to why we consider a scalar field.
While we do not have a completely satisfactory answer to this question, 
we are motivated by the fact that a significant part of
gravitational perturbations in black hole 
backgrounds are scalar in nature \cite{chandra}. 
So the scalar field does not have to be introduced {\it by hand}
from outside. 

This article is organised as follows: 
in the next section, we briefly review entanglement in
quantum mechanics and its role in entropy. In Section 3,
we set up the formalism for computing entropy of 
entangled scalar fields in flat spacetime, when the 
field is in its ground state.
In Section 4, we extend our results to 
generic coherent states and a class of squeezed states. 
In Section 5, we examine excited states and their 
effects on entanglement entropy.
In Section 6, we try to pin-point the location of the DOF
which are responsible for entanglement entropy.
In Section 7, we extend our results to 
black hole spacetimes. In Section 8, 
we conclude with a summary and open questions.

\section{Entanglement entropy} 

Consider a quantum mechanical 
system, which can be decomposed into two subsystems 
$u$ and $v$, as shown below
\footnote{for details, we refer the reader to the review
\cite{timmermans}. }:

\begin{center}

\begin{tabular}{|p{1.1cm}|}
\hline{System} \\
\hline
\end{tabular}
$=$ 
\begin{tabular}{|p{2.2cm}|}
\hline
{
 Subsystem $u$ } \\
\hline
\end{tabular}
+
\begin{tabular}{|p{2.1cm}|}
\hline
{
 Subsystem $v$} \\
\hline
\end{tabular}

\end{center}

\no
Correspondingly, the Hilbert space of the system is a Kronecker
product of the subsystem Hilbert spaces:
\be
{\cal H} = {\cal H}_u \otimes {\cal H}_v~.
\ee
If $|u_i\rangle$ and $|v_j\rangle$ are eigenbases in 
${\cal H}_u$ and ${\cal H}_v$ respectively, then 
$|u_i\rangle \otimes |v_j\rangle$ form an eigenbasis 
in $\cal H$, in terms of which a generic wavefunction 
$|\Psi\rangle$ in ${\cal H}$ can be expanded:   
\be
|\Psi \rangle 
= \sum_{ij} d_{ij} |u_i \rangle \otimes |v_j\rangle 
\in {\mathcal H} ~.
\ee
Note however, that in general, $|\Psi\rangle$ cannot be factorised  
into two wavefunctions, one in $u$ and the other in $v$: 
\be
|\Psi \rangle \neq |\Psi_u\rangle \otimes |\Psi _v \rangle ~.
\ee
Such states are called {\it Entangled States} or {\it EPR states}. 
The ones that can be factorised, on the other hand, are called
{\it Unentangled States}. 
For example, for a system with two spins, the following is 
an unentangled state: 
\be
|\uparrow \downarrow \rangle + |\uparrow \uparrow \rangle 
= |\uparrow\rangle \otimes (|\downarrow \rangle + |\uparrow \rangle )~, 
\ee
while the following is an entangled state: 
\be
|\downarrow \downarrow \rangle + |\uparrow \uparrow \rangle 
\neq |\dots \rangle \otimes | \dots \rangle~.  
\ee
Entangled states have a variety of uses, including 
in {\it Quantum Teleportation}.

Next, let us define density matrices. If the quantum-mechanical 
wavefunction of a system is known (however complex the system might be),
its density matrix is defined as: 
\be
\rho \equiv |\Psi \rangle \langle \Psi | ~. 
\ee
It can easily be verified that $\rho$ satisfies the 
following properties: 
\be
|\rho| \geq 0~,~~\rho^\dagger = \rho,
~{\rho^2 = \rho}~.
\ee
The last property is known as {\it idempotency}, from which it follows that
the eigenvalue $p_n$ of the density matrix
can only be $0$ or $1$. Thus the 
{\it Entanglement Entropy} or {\it Von Neumann Entropy}, 
defined as follows, vanishes :  
\be
S \equiv - Tr\le( \r \ln \r \ri) = - \sum_n p_n \ln p_n = 0~. 
\ee
Now, one can take the trace of $\rho$, only in the subsystem $v$,
to find the {\it Reduced Density Matrix}, which is still an
operator in subsystem $u$: 
\be
\rho_u = Tr_v(\rho) = \sum_l \langle v_l | \r | v_l\rangle 
=\sum_{i,k,j} d_{ij} d^\star_{kj} |u_i \rangle \langle u_k| ~.
\ee
For $\rho_u$, it can be shown that the following properties hold:
\be
|\r_u| \geq 0~,~~ \r_u^\dagger = \r_u~,~~
\r_u^2 \neq \r_u ~.
\ee
That is, idempotency no longer holds. As a result, its eigenvalues
now satisfy: $ 0< p_{n(u)} < 1 $, and the entanglement entropy is non-zero:
\be
S_u \equiv - Tr_u\le( \r_u \ln \r_u \ri) 
= - \sum_n p_{n(u)} \ln p_{n(u)} > 0 ~.
\ee
The ignorance resulting from tracing over 
one part of the system manifests itself as entropy. 
One important property of reduced density matrices is
that if we traced over $u$ instead and found the 
reduced density matrix $\rho_v$, then the latter
would have the same set of non-zero eigenvalues. 
Consequently, the entanglement entropies are equal, being a 
common property of the entangled system:
$S_v = S_u .$

\section{Entanglement entropy of scalar fields}

Now consider a free scalar field in $(3+1)$-dimensional flat space, with the
following Hamiltonian: 
\be
H = \f{1}{2} \int d^3x \le[ \p^2(x) + |\vec\nabla\varphi(\vec x) |^2\ri]~.
\ee
Decomposing the field and its conjugate momentum into partial waves
\ba
\varphi(\vec r)  = \sum_{lm} \f{\varphi_{lm}(r)}{r}~Y_{lm} (\theta,\phi)~,~ 
~\pi(\vec r)  = \sum_{lm} \f{\pi_{lm}(r)}{r}~Y_{lm} (\theta,\phi) 
\ea
yields:
\be
H = \sum_{lm} H_{lm}
= \sum_{lm} \f{1}{2} \int_0^\infty dr
\le\{ \p_{lm}^2(r) 
+ x^2 \le[ \f{\pa}{\pa r} \le( \f{\varphi_{lm} (r)}{r}\ri) \ri]^2 
+ \f{l(l+1)}{r^2}~\varphi_{lm}^2(r) \ri\} ~. 
\la{ham2}
\ee
Next, discretise along the radial direction with lattice spacing $a$,
such that $r \rightarrow r_i = ia;~r_{i+1}-r_i=a$. 
The lattice is terminated at a large
but finite $N$ (`size of the universe'), and an intermediate
point $n$ is chosen, which we call the {\it horizon},
and which separates the lattice points between an `inside' and
an `outside'. 
As of now, there is nothing intrinsically special about this point,
although later we will associate it with a black hole horizon. 
The discretisation is depicted below: 
\begin{center}
$| \ub{\ub{\bullet}_{1} 
 \ub{\bullet}_{2}
- \bullet 
- \bullet 
- \ub{\bullet}_{n (\mbox{Horizon})}}_{\mbox{Inside}} 
- \ub{\bullet 
- \bullet 
- \ub{\bullet}_{j} 
- \bullet 
- \bullet 
- \ub{\bullet}_{N} }_{\mbox{Outside}}
|
$
\end{center}
The discretised version of Hamiltonian (\ref{ham2}) takes the form:
\be
H_{lm}
=
\f{1}{2a}
\sum_{j=1}^N
\le[
\p_{lm,j}^2 + \le( j + \f{1}{2} \ri)^2 
\le( \f{\varphi_{lm,j}}{j} - \f{\varphi_{lm,j+1}}{j+1} \ri)^2
+ \f{l(l+1)}{j^2} 
\varphi_{lm,j}^2 
\ri]~.
\la{ham2a}
\ee 
%
Note that it resembles the Hamiltonian of a
set of $N$ coupled harmonic oscillators, with the interaction 
between them 
contained in the off-diagonal elements of the matrix $K_{ij}$
(the coordinates $x_i$ replace the field variables $\varphi_{lm}$):  
\be
H = \f{1}{2} 
\sum_{i=1}^N p_i^2 
+\f{1}{2} \sum_{i,j=1}^N x_i K_{ij} x_j 
\la{ham3}
\ee
The density Matrix, tracing over the first $n$ of $N$ oscillators
($x\equiv x_{n+1},\dots, x_N$), is given by: 
\ba
\rho \le( x; x'\ri) = 
\int \prod_{i=1}^n~dx_i~\psi (x_1,\dots,x_n;x_{n+1},\dots,x_{N})~ 
\psi^\star (x_1,\dots,x_n;x_{n+1},\dots,x_{N}') 
\la{den2} 
\ea
where ($\underbar x = Ux~,~UKU^T=~$Diagonal). The wavefunction 
\be
\psi (x_1,\dots,x_N) = 
\prod_{i=1}^N N_i 
~H_{\n_i} \le( k_{D~i}^{\f{1}{4}}~{\underbar x}_i \ri)
\exp\le( -\f{1}{2} k_{D~i}^{\f{1}{2}}~{\underbar x}_i^2 \ri) 
\la{wavefn2}
\ee
denotes the most general eigenstate of Hamiltonian (\ref{ham3}),
being a product of $N$ harmonic oscillator wavefunctions. 
Note that $\r_{out}^2 \neq \r_{out}$, since $\r_{out}$ is
mixed, although the full state is pure. Consequently, the 
entanglement entropy $= - Tr\le( \rho \ln\rho \ri) >0~. $ 
For the scalar field Hamiltonian (\ref{ham2a}), $K_{ij}$ can be read-off:
\ba
K_{ij} &=& { \frac{1}{i^2} 
\le[l(l+1) \delta_{ij} + 
\f{9}{4}~\d_{i1} \d_{j1} 
+ \le( N - \f{1}{2}\ri)^2 \d_{iN} \d_{jN}
+ \le( 
\le(i+\f{1}{2}\ri)^2 + \le(i - \f{1}{2} \ri)^2
\ri)
\d_{i,j(i\neq 1,N)} 
\ri] }  \nn\\
&&  
{
-\le[\f{(j+\f{1}{2} )^2}{j(j+1)} 
\ri] \delta_{i,j+1}
-
\le[
\f{(i+\f{1}{2} )^2}{i(i+1)} 
\ri] \delta_{i,j-1} } ~.
\la{kij1}
\ea
Note that the last two terms denote (nearest-neighbour) interaction
and originates in the derivative term in (\ref{ham2}). 
Schematically, 
{\small
\ba
K  = \le( \begin{array}{lllllll} 
{\times} & {\times} & {}& {}& {} & {} & {} \\
{\times} & {\times} & {\times} & {}& {} & {} & {} \\
{}  & {\times} & {\times} & {\times} & {} & {} & {} \\
{} & {}  & {\times} & {\times} & {\times} & {} & {} \\
{} & {} & {}  & {\times} & {\times} & {\times} & {}  \\
{} & {} & {} & {}  & {\times} & {\times} & {\times}  \\
{} & {} & {} & {} & {}  & {\times} & {\times}  
\la{mat1} \\
\end{array} \ri) 
\ea
}
where the off-diagonal terms represent interactions.
The GS of the above Hamiltonian is given by:
%
%
%
%
\be
\psi(x_1,\dots,x_N) = \prod_{i=1}^N
N_i 
\exp( -\f{1}{2} k_{Di}^{\frac{1}{2}} {\underbar x}_i^2) 
\la{gs2}
\ee
which follows from (\ref{wavefn2}), by setting all $\nu_i $ to zero.
The corresponding density matrix (\ref{den2}) can be evaluated 
exactly (the suffix $0$ signifies GS): 
\be
\r_0 (x,x') \sim \exp\le[ -(x^T \gamma x + x'^T\gamma x')/2 + x^T\beta x \ri] 
\la{gsden}
\ee
where:

\be 
\O \sim K^{1/2} = \le( \begin{array}{ll} 
A& B  \\ 
B^T & C \\
\end{array} \ri)~,~\beta=\f{1}{2}B^TA^{-1}B~,~\gamma=C-\beta~.
\ee
Note that $B$ and $\beta$ are non-zero if and only if there are
interactions. The gaussian nature of the above density matrix 
lends itself to a series of diagonalisations
\footnote{
$V\c V^T =$ diag, ${\bar\b} \equiv \c_D^{-\f{1}{2}} V \b V^T \c_D^{-\f{1}{2}}$, 
$W {\bar \b} W^T = $ diag, $v_i \in v \equiv W^T (V \c V^T)^{\f{1}{2}} VT$~. 
}
, such that it reduces to a product of $(N-n)$, $2$-oscillator
density matrices, in each of which one oscillator is traced over
\cite{sred}: 
\be
\rho_{_0}(x,x')  \sim
\prod_{i=1}^{N-n} 
\exp\le[-\f{v_i^2+v_i'^2}{2} + {\bar \beta}_i v_i v_i' \ri] ~.
\ee
The corresponding entropy is given by:
\be
S = \sum_{i=1}^{N-n} 
\le( - \ln[1-\xi_i] - \f{\xi_i}{1-\xi_i}\ln\xi_i  \ri) ~
\le[
\xi_i = \f{{\bar \b}_i}{1+ \sqrt{1-{\bar \beta}_i^2}} \ri]~.
\ee
Thus, for the full Hamiltonian $H=\sum_{lm}H_{lm}$, the entropy is:
\be
S = \sum_{l=0}^{\infty} (2 l+ 1) S_l 
\ee
where the degeneracy factor $(2l+1)$ follows from spherical symmetry
of the Hamiltonian. In practice, we will replace the upper bound of the
sum in the above to a large value $l_{max}$. 
For the interaction matrix (\ref{kij1}), the above
entropy, computed numerically, turned out to be \cite{bkls,sred}:
\be
S = 0.3 (n+1/2)^2 \equiv  
0.3 \le(\f{R}{a}\ri)^2~.
\ee
The logarithm of the entropy versus the logarithm 
of the radius of the horizon is plotted in Figure 1, and
the slope is seen to be $2$.
\begin{figure}[!ht]
\begin{center}
\epsfxsize 2.80 in
\epsfbox{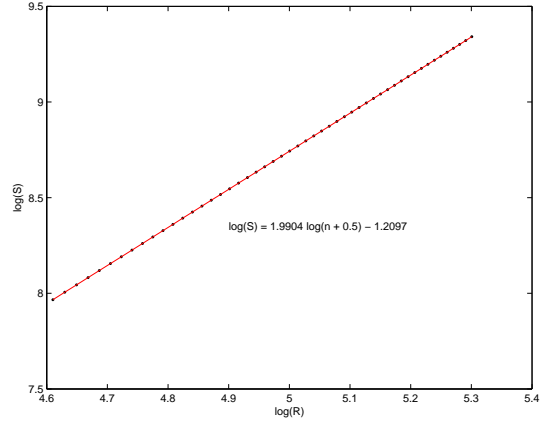}
\caption{$\ln(S)$ vs $\ln(R)$ for GS. The slope is $2$.}
\end{center}
\end{figure}

\section{Coherent and Squeezed States} 
\subsection{Coherent state}

In this section, we first examine the situation when 
GS in (\ref{gs2}) is replaced by coherent states. 
A CS (`shifted' ground state) is defined as the following,
for which $\D p \D x = \hbar/2$:
\ba
\psi(x_1,\dots,x_N)_{CS}   
&{=}& 
\prod_{i=1}^N N_i 
\exp( -\f{1}{2} k_{Di}^{\frac{1}{2}} ({\underbar x}_i - \alpha_i)^2) \\
&=& \prod_{i=1}^N
\exp(-i\underbar {p}_i \a_i)
N_i \exp( -\f{1}{2} k_{Di}^{\frac{1}{2}} {\underbar x}_i^2) 
\ea
where $\underbar p_i=-i\pa/\pa{\underbar  x_i}$~. Next, 
defining $ {\tilde x } \equiv  x - U^{-1}\a~,~~d\tilde x = dx $,
it is easy to show that the density Matrix for the CS can be written as:
\ba
\rho_{CS} \le( x; x'\ri) 
= 
\int \prod_{i=1}^n~dx_i~\psi_{CS} (x_1,\dots;x_{n+1},\dots )~
\psi_{CS}^\star (x_1,\dots;x_{n+1},\dots ) =
\rho \le( \tilde x; \tilde x'\ri)~. 
\ea
Remarkably, the density matrix has the same function form as the
GS density matrix (\ref{gsden}), albeit in terms of the tilded
variables. However, this means that the eigenfunctions will be 
identical to those for the GS, as a function of the tilded 
variables. Consequently, the eigenvalues, and hence the 
entropy will be the same as those of the GS \cite{sdss}!   

\subsection{Squeezed state}

SS are also minimum uncertainty packets, for which 
for which $\D p \D x = \hbar/2$.
However, they are characterised by 
$\D p \gg 1~$ or $\D x \ll 1~$ (or vice-versa).
Here, we consider a class of SS, characterised by 
a single parameter $r$, and given by the
following wavefunction: 
\be
\psi_{_{SS}}(x_1,\dots,x_N) 
= 
r^{N/2} \prod_{i=1}^N N_i 
\exp\le[- \sum_i r \k_{Di}^{1/2} {\underbar x}_i^2\ri] ~. 
\ee
In this case, defining a new variable as:
${\tilde x }  \equiv    
\sqrt{r}~{\underbar x}~,~~d\tilde x = \sqrt{r}~
d{\underbar x} ~,
$
the density matrix becomes:
\ba
\rho_{SS} \le( x; x'\ri) 
= 
\int \prod_{i=1}^n~dx_i~\psi_{SS} (x_1,\dots;x_{n+1},\dots )~
\psi_{SS}^\star (x_1,\dots;x_{n+1},\dots ) =
\rho \le( \tilde x; \tilde x'\ri)~. 
\ea
Once again, it has the same functional form, and the 
entropy is same as that for the ground state \cite{sdss}.

\section{First excited state}

Although, ideally one would like to compute the entanglement entropy for 
the most general eigenstate (or superpositions of such eigenstates), 
given by Eq.(\ref{wavefn2}), this turned out to be a
formidable task. Instead, to examine the role of excitations which
do not form minimum uncertainty packets, we consider the following
wavefunction, which is a superposition of $N$ wavefunctions, each
of which signifies {\it exactly} one oscillator in the first excited
state
\footnote{The analysis for two harmonic oscillators was done in \cite{ads}.}.
\ba
\psi_{_{1}}(x_1 \dots x_N) 
&=& 
{
\sum_{i=1}^N 
a_i N_i H_1 \le( k_{D i}^{\f{1}{4}} {\underbar x}_i\ri) 
\exp\le[ -\f{1}{2} \sum_{j} k_{D j}^{\f{1}{2}}~{\underbar x}_j^2 \ri] }
\\
{=} & {\sqrt{2}}&  
{
\le( a^T K_D^{\f{1}{2}} {\underbar x} \ri)
 ~\psi_0\le( x_1,\dots,x_N, \ri)}
%
\ea 
Where $a^T = \le( a_1,\dots, a_N \ri)$, and normalisation requires $a^T a=1$.
The density matrix can be written as:
\ba
\rho( x; x') 
%
=  2 \int \prod_{i=1}^n dx_i 
\le[{x}'^T  \Lambda  {x}  \ri] 
 \psi_0 \le(x_i; x \ri)  \psi_0^\star \le(x_i; x'\ri)~.
\ea
This can be evaluated {\it exactly}, yielding:
\ba
\rho(x,x') = 
\le[1 -\f{1}{2} \le( 
x^T\Lambda_\c t + x'^T \Lambda_\c x' \ri) + x^T\Lambda_\b x' \ri] 
\rho_0(x,x') 
\la{esden1}
\ea
where
\ba
\Lambda =  U^T~K_D^{\f{1}{4}}~a~a^T~K_D^{\f{1}{4}}~U \equiv
\le( 
\begin{array}{cc} 
\Lambda_A & \Lambda_B \\
\Lambda_B^T & \Lambda_C 
\end{array}
\ri) 
\ea
Without further approximations, (\ref{esden1}) cannot be factorised
into $2$-particle density matrices. However, noting that 
$\rho_0$ in the above is a gaussian (given by (\ref{gsden})),
decaying virtually to zero beyond its few sigma limits, we check
whether quantities $\epsilon_{1,2}$, defined below, are 
small for any given $\Lambda$ matrix. That is, whether:
\ba
\e_1 \equiv x_{max}^T  \Lambda_{\b}  x_{max}  \ll 1~,~ 
\e_2 \equiv x_{max}^T  \Lambda_{\gamma}  x_{max}~ \ll 1 
\ea
where 
\be
x_{max}^T = \le(\f{3 (N-n)}{\sqrt{2 Tr(\c-\b)}} \ri)\le(1,1,\dots \ri)~
\ee
corresponding to $3\sigma$ limits in the argument of $\rho_0$. If both 
$\epsilon_{1,2} \ll 1$, then we can make the following approximation:  
%
%
%

%
\ba
1 -\f{1}{2} \le(x^T\Lambda_\c x + x'^T \Lambda_\c x' \ri) + x^T\Lambda_\b x' 
\approx
\exp\le[-\f{1}{2} \le(x^T\Lambda_\c x + x'^T \Lambda_\c x' \ri) 
+ x^T\Lambda_\b x' \ri]~,
\ea
with which (\ref{esden1}) can be written as:
\ba
{
\rho(x,x') 
\approx \exp\le(  -\f{1}{2} \le( x^T \gamma' x + x'^T \gamma' x' \ri)
+ x^T \beta' x'  \ri) }
~~~
~\le[\beta' \equiv \beta + {\Lambda_\b}, 
\gamma' \equiv \gamma + {\Lambda_\c}\ri]~.
\ea
The above being of the same form as the gaussian density matrix,
but with shifted parameters, can be factorised once again into two 
particle density matrices, and the associated entanglement entropy can be 
evaluated. The rest of the computation is numerical (done using 
MATLAB). We choose the following values for parameters:
$N=300,~n=100-200,~o=10-50$, where $o$ signifies the number
of non-zero entries in the vector:
$a^{T} = \le(1/\sqrt{o} \ri) (0, \cdots, 0, {1, \cdots,1}~ ).$
For the above choice, it was seen that $\epsilon_{1,2}  \leq 10^{-3}$~. 
We give the results of our computation in the form of 
relevant graphs. Figure 2 shows the logarithm of the entropy for
the GS and ES for $o=10,30,40,50$. We see that for the ES, the 
entropy scales as $A^\a$, with lesser $\a$ for higher $o$. 
It is less than unity for any $o > 0$. The AL does not seem to
hold!  
\begin{figure}[!ht]
\begin{center}
\epsfxsize 4.70 in
\epsfbox{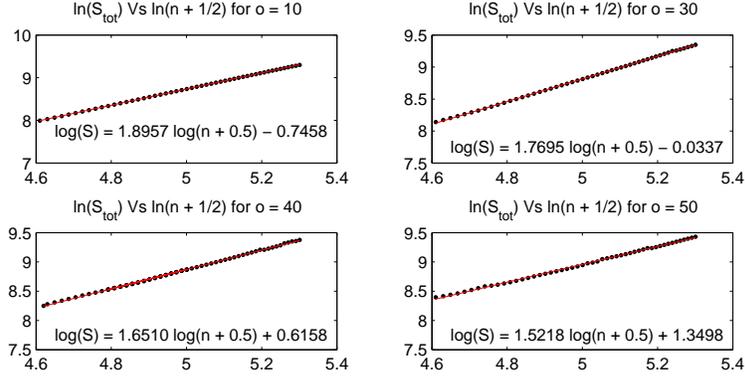}
\caption{
Plot of $\ln(S)$ vs $\ln(R)$ for ES. The slope is less than $2$.
}
\end{center}
\end{figure}
Next, in Figure 3, we plot $(2l+1)S_l$ vs $l$ for the GS as well as for
various ES and several $n$. It is seen that for the GS, there is a
peak at $l=0$ ($s$-wave), followed by another one around $l \approx 50-100$,
the degeneracy factor of $(2l+1)$ being responsible for the latter,  
For ES, the first peak shifts to $l >0$, with the shift being greater
for larger $o$. The graphs do not show a second peak, although there seems
to be an increase towards higher values of $l$. 
Thus, higher partial waves are seen to get excited
with greater excitations.

\begin{figure}[!ht]
\begin{center}
\epsfxsize 3.70 in
\epsfbox{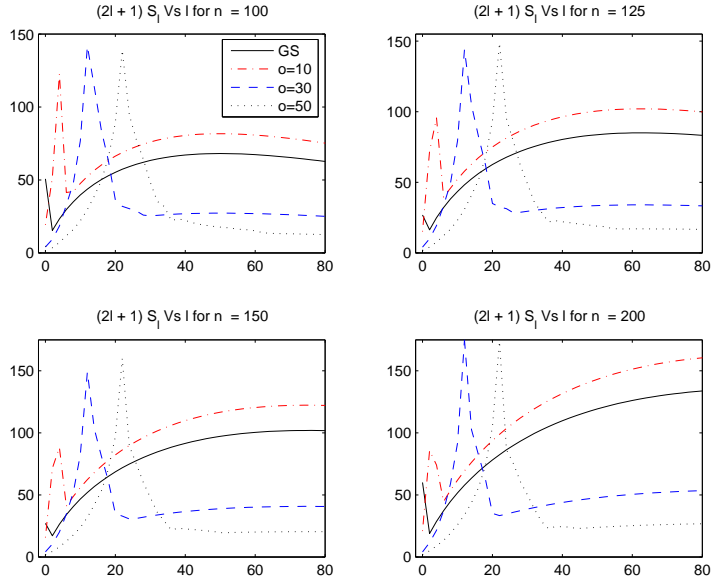}
\caption{Plot of $(2l+1)S_l$ vs $l$ for GS and different ES 
for various $n$ ($N=300$).}
\end{center}
\end{figure}

\section{Where are the degrees of freedom?} 

In this section, we address the question: where are the DOF, which
give rise to entanglement entropy for black holes?\footnote{ In
Ref. \cite{yarom2}, the authors have studied the scaling properties 
of energy fluctuations close to the boundary.}

 To this end, we re-consider the interaction matrix
$K$ in (\ref{mat1}), and set to zero, {\it by hand}, the interaction
(off-diagonal) elements except for a small window (of say size
$3\times 3$), and move the window from the origin through the horizon,
to the outer edge of the spherical box, as shown below:
%
\begin{center}
\be
K  = \le( \begin{array}{lllllll} 
{\times}  & {}  & {}& {} & {} & {} & {}  \\
{} & {\times} & {} & {} & {} & {}  & {} \\
{}  & {} & {\times} & {} & {} & {} & {}   \\
{} & {}  & {} & | \overline{{\times}} & \overline{{\times}} & \overline{{}}~~| & {}  \\
{} & {} & {}  & | {{\times}} & {{\times}} & {{\times}}| & {}  \\
{} & {} & {} & |\underline{\null} {}  & \underline{{\times}} & \underline{{\times}} |  & {}  \\
{} & {} & {} & {} & {}  & {} & {\times}  \\
\end{array} \ri) 
\ee
\end{center}
With the above modified matrix, now we compute the entanglement
entropy, and plot it as a function of the position of the centre of
the window (the green dotted boxes).  The result is shown in Figure 4
(the solid red curve). Note that when the window lies entirely outside
or inside the hrizon, the entropy is zero.  When the window just
enters the horizon, the entropy starts to rise, and when it is
symmetrically placed between the inside and outside, the entropy rises
to $100$\% of its value, and decreases again as it is further moved.
The above results confirm the intuition that entanglement between
inside and outside the horizon is the source of this entropy, and
shows how it rises from zero to its maximum value as the entangled DOF
are gradually incorporated.
\begin{figure}[!ht]
\begin{center}
\epsfxsize 2.70 in
\epsfbox{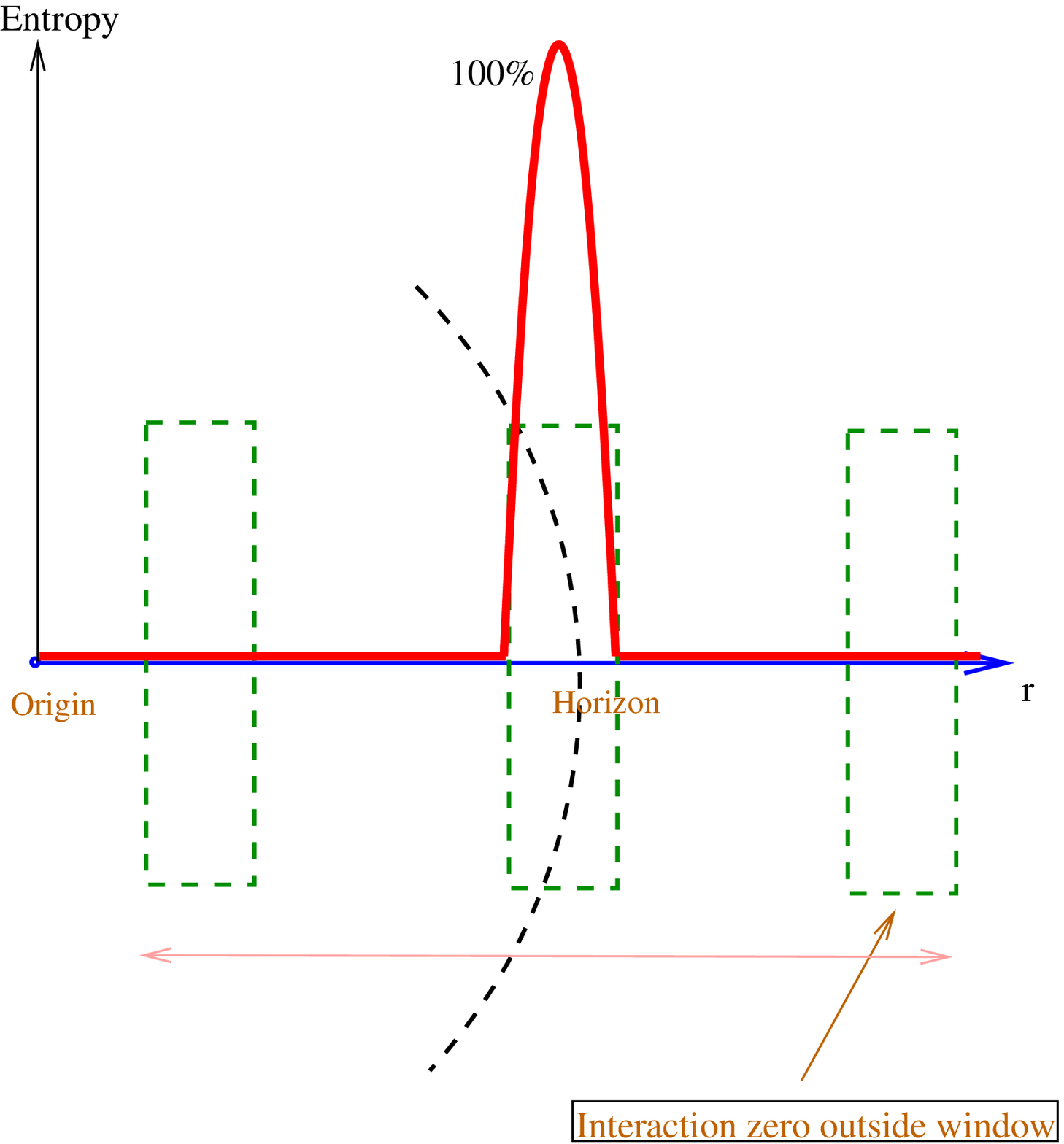}
\caption{Entanglement entropy comes from the near-horizon DOF.}
\end{center}
\end{figure}
In Figure 5, we probe this further, by keeping one extremity of the
window just outside the horizon, while increasing its width from zero
to the horizon radius. We see that with a width of one lattice
spacing, about $85\%$ of the total entropy is obtained, and within a
width of about $8-10$ lattice spacings, it reaches $100\%$. While this
shows that most of the entropy comes from DOF very close to the
horizon, a small part (about $15\%$ in this case), has its origin
deeper inside. The latter is understood to indirectly affect the DOF
near the boundary via the nearest-neighbour terms in (\ref{mat1}). We
expect a similar analysis for ES to give us a better understanding of
the scaling of the entropy with area in that case.
\begin{figure}[!ht]
\begin{center}
\epsfxsize 2.70 in
\epsfbox{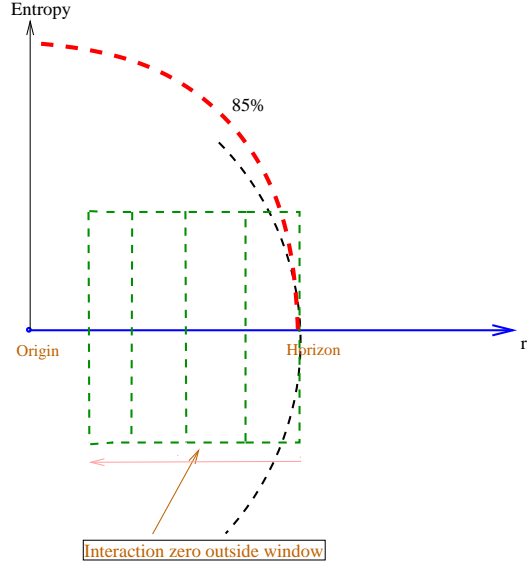}
\caption{Rise of entropy as interaction region is increased.}
\end{center}
\end{figure}

\section{Generalisation to black hole spacetimes}

In this section, we generalise the framework of our calculations,
so that they are applicable to black hole spacetimes. We start
with the Schwarzschild metric with a horizon at $r=r_0$: 
\be
ds^2 = -f(r) dt^2 + \f{dr}{f(r)} + r^2 d\O^2 ~,~~f(r)= 1 - \f{r_0}{r} 
\la{sch1}
\ee
and define the following coordinate transformation from $(t,r)$ 
to $(\tau,R)$ coordinates:
\ba
\tau = t + r_0 \le[ \ln\le( \f{1-\sqrt{r/r_0} }{1+\sqrt{r/r_0}} \ri) 
+ 2\sqrt{\f{r}{r_0}}  \ri]~,~
R = \tau + \f{2~r^{\f{3}{2}}}{3\sqrt{r_0}} ~,~
r = \le[ \f{3}{2} \le( R-\t\ri) \ri]^{\f{2}{3} } r_0^{\f{1}{3} }~,
\la{transfn1}
\ea
such that (\ref{sch1}) transforms to the following metric, in 
{\it Lemaitre coordinates}:  
\be
ds^2= -d\t^2 + 
\f{dR^2}{\le[ \f{3}{2r_0} \le( R-\t\ri) \ri]^{\f{2}{3}} }
+ \le[ \f{3}{2r_0} \le( R-\t\ri) \ri]^{\f{4}{3}}r_0^{\f{2}{3} }~d\O^2 ~. 
\ee
The Hamiltonian for a free scalar field in the above spacetime can be
written as 
\footnote{the current analysis is done for $l=0$. Generalisation
to any $l>0$ is straightforward.}:
\be
H(\t)= \f{1}{2} \int_\tau^\infty dR  
\le[
\f{2\p(\t,R)^2}{3(R-\t)} + \f{3}{2}~r~(R-\t) \le( \pa_R \phi(\t,R) \ri)^2  
\ri]~.
\la{schham}
\ee
Next, choose a {\it fixed} Lemaitre time, say $\t=0$ and perform the
following field redefinitions:
\be
\p(r) = \sqrt{r} \p_1(r) ~,~~\phi (r) = \f{\phi_1(r)}{r}~.  
\ee
Then, at that {\it fixed time}, the 
Hamiltonian (\ref{schham}) transforms to: 
\be
H(0) = \f{1}{2} \int_0^\infty dr
\le[
\p_1(r)^2 + r^2 \le( \pa_r \f{\phi_1}{r} \ri)^2
\ri]~.
\ee
Note that this is identical to the Hamiltonian in flat spacetime,
Eq.(\ref{ham2}).  
Also, it follows from Eq.(\ref{transfn1}) 
that the horizon now corresponds to 
%
$R = \f{2}{3} r_0 $.
Thus, one is free to trace over either the region
$R=0 \rightarrow \f{2}{3} r_0$ or over the region
$R= \f{2}{3} r_0 \rightarrow \infty $, and 
all our results go through for a fixed $\t$.  
Once again, the AL holds for GS, CS and SS, while it does
not for ES. We also show that although the degrees of 
freedom near the horizon contribute most to the entropy, 
the DOF deep inside must also be taken into account for the 
AL.

\section{Summary and open questions}

In this article, we have shown that entanglement between DOF
inside and outside the horizon is a viable source of 
black hole entropy. However, while the AL holds for 
the GS, CS, and SS, this entropy scales as an area power
less than unity, when the considered wavefunction is a 
superposition of first ES. Furthermore, while the DOF near the horizon
contribute most to the entropy, those farther away give a small
contribution as well. We would like to investigate this further 
for ES, as that might provide a physical understanding 
of the deviation from the AL. We would also like understand
the phhenomena of Hawking radiation and information loss in this 
icture. We hope to report on this elsewhere \cite{DS-new}.

\ack
We would like to thank M. Ahmadi, R. K. Bhaduri, C. Burgess,
A. Dasgupta, J. Gegenberg, A. Ghosh, V. Husain, G. Kunstatter, S. Nag,
A. Roy, T. Sarkar and R. Sorkin for useful discussions. 
We also thank the anonymous referees of
\cite{sdss} for several
important comments and suggestions. SS would like to thank the
Dept. of Physics, Univ. of Lethbridge for hospitality where most of
this work was done. This work was supported by the Natural
Sciences and Engineering Research Council of Canada and the 
Perimeter Institute for Theoretical Physics.

\end{document}